\documentclass[]{spie}  

\usepackage{amsmath,amsfonts,amssymb}
\usepackage{aas_macros}
\usepackage{graphicx}
\usepackage[colorlinks=true, allcolors=blue]{hyperref}
\usepackage{booktabs}

\title{The Pandora SmallSat: Target Selection and Early Science}
\author[*a]{Emily A. Gilbert}
\author[b]{Elisa V. Quintana}
\author[c]{Jessie L. Dotson}
\author[b]{Knicole D. Col\'on}
\author[b]{Thomas Barclay}
\author[d]{Pete Supsinskas}
\author[d]{Jordan Karburn}

\author[e,f,g]{D\'aniel Apai}
\author[b,h]{Christina Hedges}
\author[i]{Benjamin V. Rackham}
\author[j]{Jason F. Rowe}
\author[a]{David Ciardi}
\author[k]{Natalie H. Allen}
\author[a]{Jessie Christiansen}
\author[k,l]{N\'estor Espinoza}
\author[b, m]{Trevor O. Foote}
\author[f]{Nguyen Fuda}
\author[n]{Aylin Garcia Soto}
\author[o]{Thomas P. Greene}
\author[j]{Kelsey Hoffman}
\author[b,m]{Rae Holcomb}
\author[b,m]{Benjamin J. Hord}
\author[b,m]{Aishwarya R. Iyer}
\author[a]{Aurora Y. Kesseli}
\author[b,p]{Veselin B. Kostov}
\author[q]{Nikole K. Lewis}
\author[r]{Sarah E. Logsdon}
\author[s]{Andrew W. Mann}
\author[t]{Megan Mansfield}
\author[u]{James Mason}
\author[d]{Peter McGill}
\author[b,x]{Julie Moquin}
\author[m]{Brett M. Morris}
\author[b]{Gregory Mosby}
\author[m]{Susan E. Mullally}
\author[n]{Elisabeth R. Newton}
\author[b,t]{Joe P. Renaud}
\author[w]{Yoav Rotman}
\author[f]{Suman Saha}
\author[b]{Joshua E. Schlieder}
\author[w]{Luis Welbanks}
\author[t]{Lindsey S. Wiser}
\author[b]{Allison Youngblood}
\author[b]{Robert T. Zellem}

\affil[a]{Caltech/IPAC-NASA Exoplanet Science Institute, Pasadena, CA 91125, USA}
\affil[b]{NASA Goddard Space Flight Center, Greenbelt, MD 20771, USA}
\affil[c]{NASA Ames Research Center, Moffett Field, CA 94035, USA}
\affil[d]{Lawrence Livermore National Laboratory, Livermore, CA 94550, USA}
\affil[e]{Steward Observatory, The University of Arizona, Tucson, AZ 85721, USA}
\affil[f]{Lunar and Planetary Laboratory, The University of Arizona, Tucson, AZ 85721, USA}
\affil[g]{James C. Wyant College of Optical Sciences, The University of Arizona, AZ 85721, USA}
\affil[h]{University of Maryland, Baltimore County, Baltimore, MD 21250, USA}
\affil[i]{Department of Earth, Atmospheric and Planetary Sciences, and Kavli Institute 
for Astrophysics and Space Research, Massachusetts Institute of Technology, Cambridge, MA 02139, USA}
\affil[j]{Department of Physics and Astronomy, Bishops University, 2600 Rue College, Sherbrooke, QC J1M 1Z7, Canada }
\affil[k]{Department of Physics and Astronomy, Johns Hopkins University, Baltimore, MD 21218, USA}
\affil[l]{Space Telescope Science Institute, 3700 San Martin Drive, Baltimore, MD 21218, USA}
\affil[m]{NASA Postdoctoral Program Fellow}
\affil[n]{Department of Physics and Astronomy, Dartmouth College, Hanover NH 03755, USA}
\affil[o]{California Institute of Technology/IPAC, Pasadena, CA 91125, USA}
\affil[p]{SETI Institute, 189 Bernardo Ave, Suite 200, Mountain View, CA 94043, USA}
\affil[q]{Department of Astronomy, Cornell University, Ithaca, NY 14853, USA}
\affil[r]{NSF NOIRLab, 950 N.\ Cherry Ave., Tucson, AZ 85719, USA}
\affil[s]{Department of Physics and Astronomy, The University of North Carolina at Chapel Hill, Chapel Hill, NC 27599, USA}
\affil[t]{University of Maryland, College Park, College Park, MD 20742 USA}
\affil[u]{Johns Hopkins University Applied Physics Laboratory, Laurel, MD 20723, USA}
\affil[v]{Dartmouth College, Hanover, NH 03755, USA}
\affil[w]{School of Earth and Space Exploration, Arizona State University, 781 Terrace Mall, Tempe, AZ, 85287, USA}
\affil[x]{Southeastern Universities Research Association, Washington, DC 20005, USA}

\authorinfo{*egilbert@caltech.edu}

\begin{document} 
\maketitle

\begin{abstract}
The Pandora spacecraft is a smallsat selected in NASA’s inaugural class of Pioneers missions. The aim of the Pandora mission is to disentangle stellar activity signals from exoplanet atmospheric transmission spectroscopy measurements. Pandora accomplishes this goal with its combined visible photometry and NIR spectroscopy that observe each target in the prime mission for 10 separate transit observations, each spanning 24 hours at a time. 
Targets were carefully selected for the Pandora prime mission. Pandora is specifically studying planets around low-mass stellar hosts (K and M dwarf stars) spanning a range of rotation periods and activity levels. The planets range in size from 1.5 - 15 Earth radii and orbital periods ranging from 1.7 - 11.5 days. 
In addition to the 10 transit visits planned for each of the 20 prime mission targets, Pandora is observing a number of auxiliary targets encompassing a variety of science topics. This includes topics such as stellar activity and young systems, transit timing variations of exoplanets, brown dwarf compositions, exoplanet phase curves, etc. We will discuss the selection of these targets, how they have been scheduled to fill in gaps in prime mission operations, and early science results.
If the mission is extended, there is further opportunity for accomplishing additional science. Here we will discuss the possible options for extended mission operations and science.

\end{abstract}

\section{INTRODUCTION}
\label{sec:intro}




The field of exoplanets has evolved significantly since the discoveries of the very first planets outside of our own Solar System \cite{Wolszczan1992, Mayor1995}. Early exoplanet detections focused on individual systems and primarily measured either planet mass or radius. Over time, technology and techniques advanced, and we were able to measure both mass and radius for many known exoplanets. This enabled us to measure bulk densities for individual planets, allowing astronomers to better understand planetary compositions (namely, e.g. is this planet rocky or gaseous?) and assess planet formation models. 

Building upon the study of planetary bulk compositions, exoplanetary observations then advanced to studying the chemical make-up of planetary atmospheres. Understanding atmospheric composition provides valuable insight into how planets form and evolve, thus moving us closer to the goal of understanding which planets might be habitable. In particular, carbon and oxygen molecules in planet atmospheres are key tracers of a planet's formation and evolutionary history\cite{Mordasini2024}.

Observations from both ground- and space-based facilities have been pushing towards this goal of characterizing planet atmospheres since the discovery of the first transiting exoplanet\cite{Charbonneau2000} and the detection of its atmosphere\cite{Charbonneau2002}. These facilities have readily shown that water (and thus oxygen) is prevalent in exoplanet atmospheres. 

JWST, which launched in 2022, has significantly advanced our ability to study planet atmospheres. With JWST, we have shown that carbon species (e.g., carbon dioxide and methane) are common in exoplanet atmospheres. JWST is limited in sensitivity, and thus largely limited to studying planets orbiting low-mass stars where planetary atmospheric signals are detectable. This is challenging because low-mass stars are known to be highly magnetically active for long portions of their lifetimes. Cool spots and bright faculae on stellar surfaces can vary both spatially and temporally, presenting significant challenges when it comes to extracting planetary spectra from transmission spectroscopy observations. Starspots are also cool enough that water molecules can form, thus limiting our ability to study water in planet atmospheres, as it is challenging to determine the source of water signals. Early results from JWST\cite{Moran2023, Rackham2023, Canas2026} have been plagued with difficulties interpreting results such as these.

\begin{figure}
    \centering
    \includegraphics[width=0.99\linewidth]{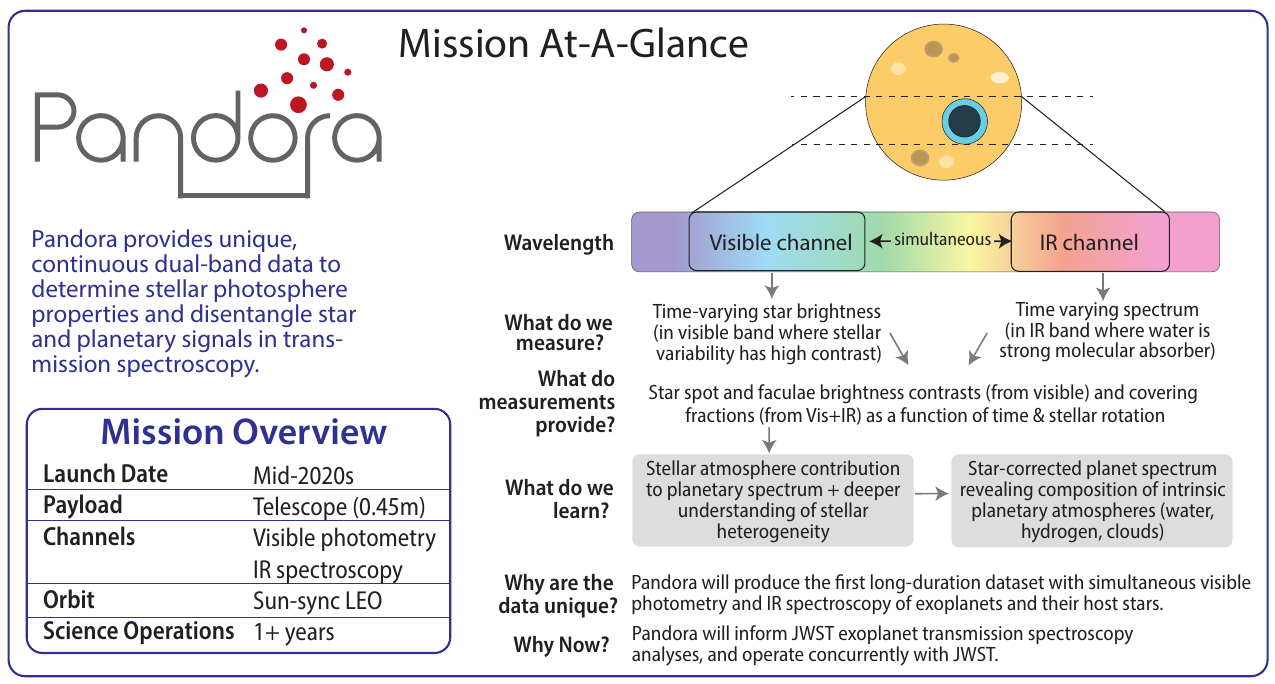}
    \caption{A summary of the Pandora mission's capabilities and goals. Pandora can uniquely perform simultaneous visible photometry and near infrared spectroscopy of individual targets on a space-based platform. In a sun-synchronous, low Earth orbit (LEO), Pandora has the ability to dwell on targets for long (days-weeks) periods of time, enabling us to probe stellar contamination over the rotation periods of Pandora's target stars.}
    \label{fig:mission-glance}
\end{figure}

Pandora is specifically designed to address the issue of stellar contamination in planetary transmission spectroscopy. Pandora is a SmallSat that was selected as a part of the inaugural class of NASA Astrophysics Pioneers missions in 2021.\cite{Quintana2024} Pandora is a partnership between NASA and Lawrence Livermore National Laboratory and includes additional partners from across the United States and Canada. Pandora is designed to observe transmission spectra of exoplanets in order to disentangle stellar activity signals from those of planet atmospheric measurements. Pandora will accomplish this goal by utilizing simultaneous NIR ($\sim$0.85-1.62 micron) spectroscopy combined with optical ($\sim$0.38-0.75 micron) photometry, see Fig. \ref{fig:mission-glance}. On January 11, 2026, Pandora launched to a Sun-synchronous, low-Earth orbit on board a Falcon 9 rocket out of Vandenberg Space Force Base, in Santa Barbara County, California. 




\section{PRIME MISSION TARGET SELECTION}
For the Prime Mission, Pandora will observe 20 separate known transiting exoplanet targets. Pandora will observe 10 transits of each of the 20 planets (including 24 hours of baseline observations) thus enabling it to achieve its prime mission goal of disentangling stellar activity signals in exoplanet transmission spectroscopy observations. To this end, we have carefully selected 20 exoplanets orbiting 19 K- and M-dwarf host stars which will be observed as a part of the Pandora Prime Mission.

Pandora targets were selected by the Science Team in order to span a variety of different planet and stellar parameters. Planets were chosen to span a range of masses, radii, compositions, and orbital periods while host stars were selected to span a range of rotation periods and sizes.

For our Prime Mission, we first filtered candidate targets for targets that met the following criteria in order to be amenable to transmission spectroscopy observations with Pandora:

\begin{itemize}
\setlength\itemsep{0pt}
    \item H$_{mag}<$11 and 7$<$J$_{mag}<$11.5
    \item T$_{eff} <$ 5300 K
    \item Planet orbital period $<$ 18 days 
    \item Planets are listed as “Confirmed” on the NASA Exoplanet Archive (i.e. not candidates)
\end{itemize}

\noindent\textbf{Magnitude requirement -} This requirement ensures that the targets are sufficiently bright to meet our signal-to-noise ratio (SNR) requirements to achieve our science goals, but not so bright that they saturate the detector.

\noindent\textbf{Temperature requirement -} We selected for cooler, Main Sequence stars, which helps to ensure that the planet transit signals are detectable as the signal size is dependent on the stellar radius. Furthermore, cooler stars are known to be more active and are potentially cool enough to have molecules like water form in their atmospheres. These features are the type of stellar activity that Pandora is trying to mitigate, so we selected stars where we know this activity will occur.

\noindent\textbf{Orbital period requirement -} We limited planetary orbital periods to $<$18 days in order to make sure observations can be scheduled within the Pandora Prime Mission. There need to be a minimum of 10 transits within the year-long mission, and we are not able to observe every single transit due to scheduling and observability constraints. Limiting orbital periods to 18 days is a conservative cutoff to ensure we can carry out all Prime Mission observations.

\noindent\textbf{Confirmed planet requirement -} Pandora Prime Mission targets were selected from planets with ``confirmed" status in the NASA Exoplanet Archive \cite{christiansen2025} in order to avoid observing targets that may turn out to be false positives. Exoplanets in the NASA Exoplanet Archive include planets that are statistically validated, but do not have measured masses. In the event that there were no published masses, we used mass/radius relationships\cite{Chen2017} and/or private communications in order to obtain unpublished masses for all planets in our target list.

After filtering for these selection criteria, we then calculated the Transmission Spectroscopy Metric (TSM)\cite{Kempton2018} for each target. TSM is a standard metric for assessing the feasibility of studying the atmospheres of transiting exoplanets by taking into account the size and mass of the planet, the radius of the host star, the brightness of the host star, and the planet’s equilibrium temperature. We selected long list targets with TSM $>$ 1 as an initial starting point to ensure that atmospheric measurements were at all feasible, then custom filtered further in order to meet Pandora's specific needs (see below).

From here, the Pandora Stellar Contamination Science Working Group team developed an additional metric -- Signal Detectability Metric (SDM) -- that was used to prioritize targets for study with Pandora. The SDM builds upon the TSM by also taking into account stellar activity using data from the Transiting Exoplanet Survey Satellite (TESS)\cite{Ricker2015} mission in order to assess rotational brightness modulation. The SDM is defined as:

\begin{equation}
    SDM = TSM * (1+ w *A)
\end{equation}

where TSM is the Transmission Spectroscopy Metric, w is a weighting factor, and A is the peak-to-peak amplitude of stellar variability. The weighting factor is set such that, for a value of w = 100, w doubles the SDM for a star with A = 1$\%$. 

\begin{figure}
    \centering
    \includegraphics[width=0.99\linewidth]{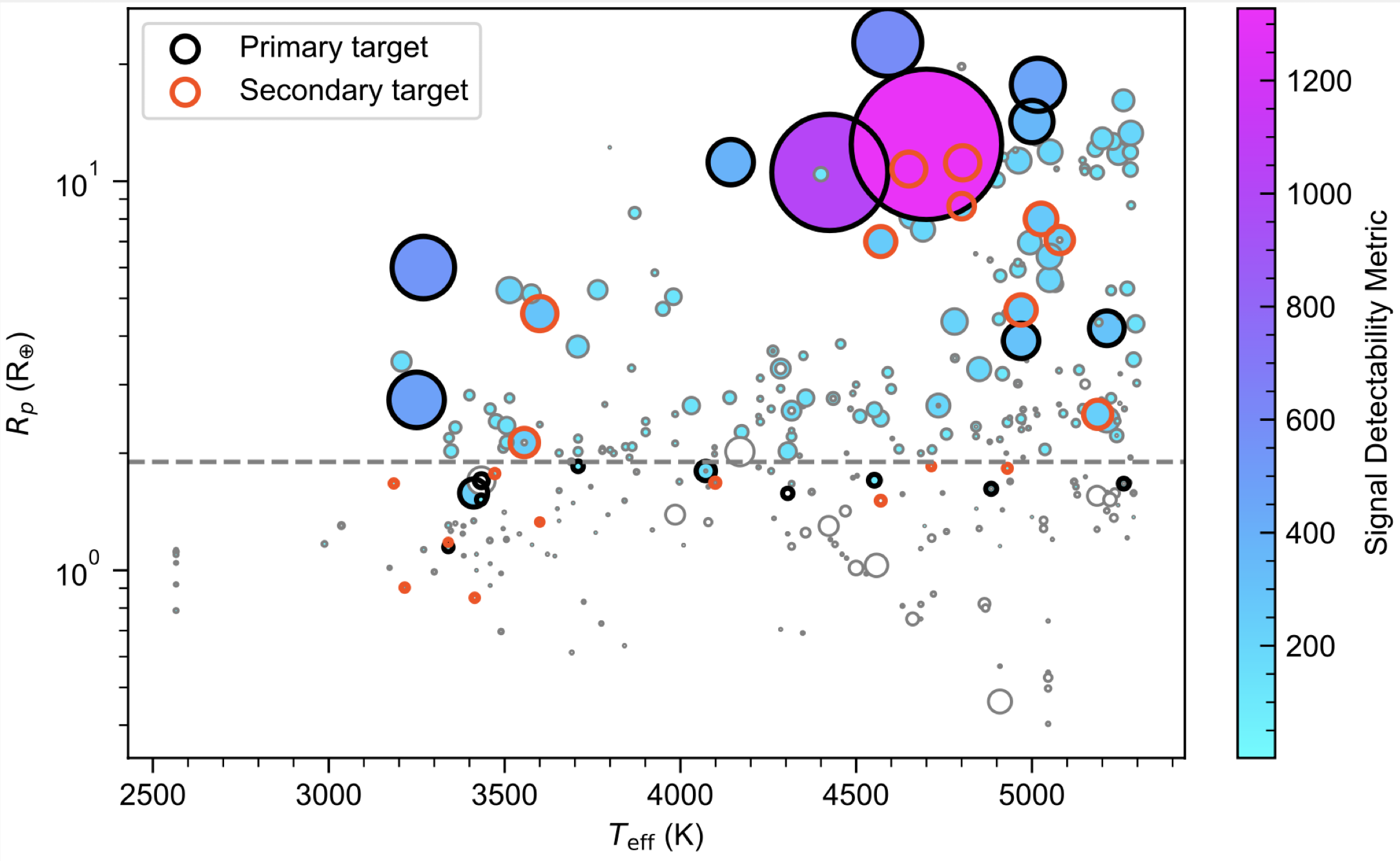}
    \caption{In order to select our final Prime Mission targets, we plotted each target in our long list to show the range of planet radius and stellar host effective temperatures covered. We selected the top 10 gaseous (above 1.9 R$_\oplus$) and top 10 terrestrial (below 1.9 R$_\oplus$) targets to be our 20 Prime Mission targets, as indicated by the black ringed points. An additional 20 targets were selected as Secondary Targets as backups to the Prime Mission targets, shown with red rings.}
    \label{fig:targets}
\end{figure}

Then, by taking into account the SDM, we selected our final targets by carefully balancing across spectral type of host stars (late G to M spectral classes) and planet sizes (Earth- to Jupiter-size). We plotted all of our long list targets on a figure of planet radii as a function of host star effective temperature (see Fig. \ref{fig:targets}). We then drew a line at 1.9 R$_\oplus$ to separate between terrestrial and gaseous planets. From here, we selected the top 10 planets by SDM on either side of the line in order to compile our primary target list. We also selected a secondary target list of 20 additional targets to be used in case the primary 20 did not meet our needs. Together, the primary and secondary lists comprise our ``Top 40" target list.

\subsection{Target Feasibility}
Pandora's science team evaluated each of the candidate targets not only on signal detectability, but also on the impact Pandora data would make scientifically and the observability of each target with the observatory. This final filtering distilled the longer top 40 targets to the top 20 targets currently on Pandora's prime mission target list.

\subsubsection{Scientific Impact}
Candidate targets were evaluated by the Science Team on the magnitude of the scientific impact that Pandora data would have on the understanding of each planet and system. This included a literature deep dive into the current understanding of each planet's atmospheric composition as well as assessing what data in which wavelength regimes were currently available for each candidate target. Targets where Pandora's wavelength coverage or unique ability to capture simultaneous visible and near infrared observations over long timescales were prioritized. This also included targets with a publication history that explicitly finds evidence of stellar activity contaminating the planetary spectrum or in which water is likely to be present but has only been ambiguously detected due to poor understanding of the host star. Targets that already had significant resources devoted to disentangling stellar activity from the exoplanet spectrum or did not display atmospheric features were deprioritized, as it is unlikely that Pandora data would have as large of an impact in these cases.

\subsubsection{Spectral Overlap}
\begin{figure}
    \centering
    \includegraphics[width=0.99\linewidth]{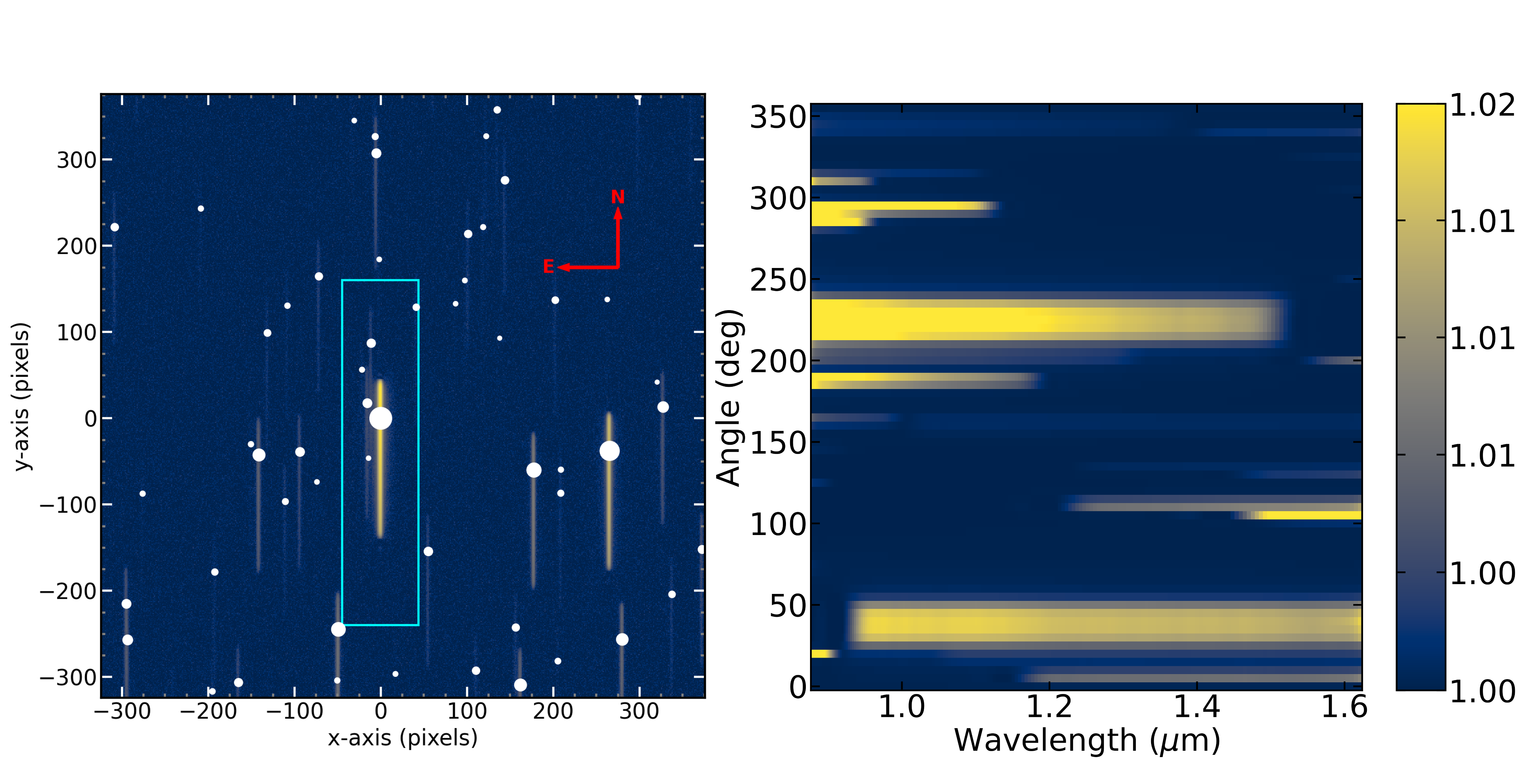}
    \caption{An example of our spectral overlap analysis for GJ 3090. \textit{Left:} A simulate starfield with synthetic Pandora spectral traces in realistic brightnesses and orientations for the star field. The size of Pandora's NIRDA detector is denoted by the light blue box. \textit{Right:} A representation of the spectral overlap when the star field is rotated to every possible orientation. Colors represent amount of total flux at different wavelengths in the trace compared to the amount of flux from the target trace.}
    \label{fig:trace_sim}
\end{figure}

Pandora's optical system contains a dispersive element in the optical path of the near infrared channel, producing a dispersed spectral trace on the near infrared detector, similar to that of JWST's NIRSpec. Because of this, observation planning must necessarily consider overlap of spectral traces from nearby targets on the spectral trace of the target star. To this end, we performed simulations of the spectral traces of all top 40 targets using fluxes and effective temperatures provided by Gaia DR3\cite{Vallenari2023} to generate synthetic stellar spectra using PHOENIX\cite{Husser2013}. These synthetic spectra were then passed through the core logic of the Pandora observation simulation package \texttt{pandorasim} to generate simulated on-detector traces. Since overlap is a strong function of star field orientation, the full range of spacecraft rolls were simulated for each target, noting which parts of the target spectral trace overlapped with a background trace and to what level (Fig \ref{fig:trace_sim}).

Targets with significant spectral overlap at the 1$\%$ level across more than half of spacecraft rolls were discarded from the prime mission target list. This also included all of the binary systems on the top 40 target list. While it is not impossible to disentangle the target spectral trace from an overlapping background trace, the choice to remove targets with significant spectral overlap was motivated by the need to greatly simplify the data analysis efforts in a low cost cap environment and to ensure any spectral changes observed were truly from the target star rather than erroneous subtraction of an overlapping trace.

\subsubsection{Target Visibility and Scheduling}
We developed custom software (\texttt{pandoravisibility}) to calculate the visibility of an arbitrary set of astronomical coordinates to Pandora. Using the spacecraft ephemeris in the form of Pandora's Two Line Element (TLE), the software calculates the vector between the spacecraft, the target, the Sun, the Earth, and the Moon to check that the vector between the spacecraft and the target does not violate any Sun, Earth, or Moon keep out angles at each user-defined time step. Additionally, the software tracks the angles between Pandora's two star trackers and the Sun, Earth, and Moon to ensure at least one star tracker has a clear field of view at all times to maintain pointing stability. Because the validity of star tracker pointing is heavily tied to the roll of the spacecraft, \texttt{pandoravisibility} also determines the optimum spacecraft roll that maximizes the duration of star tracker visibility while maintaining sufficient sunlight on the spacecraft's solar panels to ensure power-positivity.

This visibility calculation is used to determine the windows of visibility across Pandora's Prime Mission in which each target can be scheduled and is included in the custom scheduling software, \texttt{pandorascheduler}\cite{Foote2023}. This software is used to generate an observing calendar on the order of months, first prioritizing the scheduling of Prime Mission targets centered on their transits, before scheduling the remaining time with auxiliary targets provided by the science team for scientific investigations beyond the scope of Pandora's Prime Mission objectives. Included in these auxiliary targets are targets scheduled during the occultation of Prime Mission targets so that the observatory always has a science target to look at, greatly enhancing the observatory's duty cycle. This long term calendar is then run through a smaller version of the scheduling software to apply ephemeris updates to week-long calendars throughout the mission in order to keep observation timing updated as Pandora's orbit evolves.

\begin{table}[ht]
\centering
\small
\caption{Pandora Science Targets (as of July 30, 2025). Missing values indicated by ``\ldots''.}
\label{tab:pandora_targets}
\begin{tabular}{llrrrlrrrr}
\toprule
Star & Planet & RA & Dec & $V_\mathrm{mag}$ & $J_\mathrm{mag}$ & Sp.\ Type & $P_\mathrm{rot}$ [d] & $R_p$ [$R_\oplus$] & Transit Depth [\%] \\
\midrule
GJ 3090  & b & 20.4384  & $-46.7147$ & 11.403 & 8.168  & M2   & \ldots & 2.13  & 0.14  \\
GJ 3470  & b & 119.774  & 15.3912    & 12.332 & 8.794  & M1.5 & \ldots & 4.57  & 0.52  \\
HAT-P-11 & b & 297.71   & 48.0819    & 9.46   & 7.608  & K    & \ldots & 4.9   & 0.33  \\
HAT-P-12 & b & 209.389  & 43.4933    & 12.665 & 10.794 & K4   & \ldots & 10.75 & 1.977 \\
HAT-P-18 & b & 256.346  & 33.0123    & 12.597 & 10.822 & K2   & \ldots & 11.15 & 2.179 \\
HAT-P-26 & b & 213.157  & 4.05942    & 11.757 & 10.08  & G    & \ldots & 7.06  & 0.54  \\
HD 73583 & b & 129.688  & -13.2565 & 9.67   & 7.649  & K4   & 12.2   & 2.79  & \ldots \\
L 98-59  & c & 124.533  & -68.3145 & 11.685 & 7.933  & M3   & 80.9   & 1.39  & 0.166 \\
TOI-2076 & b & 217.392  & 39.7904    & 9.139  & 7.613  & G    & 7.21   & 2.39  & 0.105 \\
TOI-244  & b & 10.5698  & -36.718  & 12.861 & 8.827  & M2.5 & 53.3   & 1.52  & 0.106 \\
TOI-270  & d & 68.4161  & -51.9574 & 12.603 & 9.099  & M3   & \ldots & 2.13  & 0.09  \\
TOI-3629 & b & 359.793  & 39.3143    & 14.639 & 11.424 & M1   & \ldots & 8.29  & \ldots \\
TOI-3884 & b & 181.572  & 12.507     & 15.744 & 11.127 & M    & \ldots & 6.43  & 3.074 \\
TOI-674  & b & 164.587  & -36.8581 & 14.203 & 10.359 & M2   & 52.0   & 5.25  & 1.11  \\
TOI-942  & b & 76.6496  & -20.2456 & 11.982 & 10.231 & K2.5 & 3.39   & 3.89  & 0.148 \\
TOI-942  & c & 76.6496  & -20.2456 & 11.982 & 10.231 & K2.5 & 3.39   & 4.67  & 0.229 \\
WASP-107 & b & 188.386  & -10.1462 & 11.592 & 9.378  & K6   & 17.0   & 10.4  & 2.091 \\
WASP-52  & b & 348.495  & 8.76108    & 12.192 & 10.588 & K2   & \ldots & 14.24 & 2.71  \\
WASP-69  & b & 315.026  & -5.09486 & 9.873  & 8.032  & K5   & 23.07  & 12.44 & 1.8   \\
WASP-80  & b & 303.167  & -2.14444 & 11.841 & 9.218  & K7   & \ldots & 11.2  & 2.937 \\
\bottomrule
\end{tabular}
\end{table}

\section{AUXILIARY SCIENCE}
In the year-long Prime Mission, there is some amount of time when Pandora will not be observing Prime Mission candidates, taking calibration data, or downlinking data when additional data may be taken. This occurs when there is no primary target available due to observational limitations (such as Earth occultation) or there is no Prime Mission target transit occurring. During this time, we are planning to conduct auxiliary science observations to maximize the scientific return of the spacecraft. Pandora's simultaneous visible photometry and NIR spectroscopy enables a variety of additional science cases. Pandora's fast slew times ($<$ 2 mins) and flexible readout cadences (0.2 sec intervals) can support a variety of observing modes. Frequent Earth occultations as a result of being in low Earth orbit also provide ample opportunities for such observations to occur.

Pandora can build on its existing observational plan by conducting more transit observations to study planet atmospheres in greater detail. Pandora can observe exoplanet secondary eclipses and/or full phase curves or study atmospheric escape via the helium triplet by targeting targets with shorter orbital periods than those currently targeted by the mission. Other exoplanet science topics that Pandora can contribute to include: precise timing measurements for transit timing variation studies, planet searches, or follow-up/confirmation of planet candidates from missions like TESS. 

The spacecraft can provide useful observations for stellar astrophysics such as stellar activity monitoring of spots and flares, spectroscopic studies of flare properties, and asteroseismology. Pandora can also study brown dwarfs, measuring their rotation periods and obtaining spectra. Solar System science is another topic where Pandora can contribute meaningful observations. Pandora can observe Solar System occultations, which occur when one object passes in front of another, providing a unique opportunity to study the sizes, shapes, and atmospheres of these objects. We can also conduct comet and asteroid monitoring of near-Earth objects. Finally, Pandora is capable of conducting a wide range of astrophysical observations as well. In particular, Pandora can follow up supernovae, gravitational waves, and other transients, which is particularly useful with the recent first light of the Rubin Observatory. With Pandora's simultaneous visible photometry and NIR spectroscopy, Pandora is a valuable facility for any number of science cases. It will provide exciting observations during its Prime Mission and will continue to conduct observations in the event of an extended mission as well.




\section{SUMMARY}

The Pandora mission aims to characterize exoplanet atmospheres and their host stars. In order to achieve this goal, we have selected 20 Prime Mission targets that meet a number of criteria  in order to produce the most meaningful science results possible. We have also selected 20 Secondary Targets to be used as backups, as well as Auxiliary Targets to fill any additional time available during the Prime Mission. For the most up to date information on the status of the Pandora mission and the target list, please see \url{https://pandorasat.com/targets/}.

\acknowledgments 
Pandora is supported by NASA's Astrophysics Pioneers Program. Pre-proposal work was supported by GSFC's Internal Research and Development Program and a Mission Planning Lab at Wallops Flight Facility. We are indebted to the many people who contributed to the various iterations of Pandora, and to the technical team at LLNL who have been key to the success of the mission.

We are grateful for all the contributions that many people have provided during the development of the Pandora SmallSat.

This research has made use of the NASA Exoplanet Archive, which is operated by the California Institute of Technology, under contract with the National Aeronautics and Space Administration under the Exoplanet Exploration Program.

\bibliography{pandora} 

@ARTICLE{Husser2013,
   author = {{Husser}, T.-O. and {Wende-von Berg}, S. and {Dreizler}, S. and 
	{Homeier}, D. and {Reiners}, A. and {Barman}, T. and {Hauschildt}, P.~H.
	},
    title = "{A new extensive library of PHOENIX stellar atmospheres and synthetic spectra}",
  journal = {\aap},
archivePrefix = "arXiv",
   eprint = {1303.5632},
 primaryClass = "astro-ph.SR",
     year = 2013,
    month = may,
   volume = 553,
      eid = {A6},
    pages = {A6},
      doi = {10.1051/0004-6361/201219058},
   adsurl = {http://adsabs.harvard.edu/abs/2013A%26A...553A...6H}
}

@ARTICLE{Kempton2018,
       author = {{Kempton}, Eliza M. -R. and {Bean}, Jacob L. and {Louie}, Dana R. and
         {Deming}, Drake and {Koll}, Daniel D.~B. and {Mansfield}, Megan and
         {Christiansen}, Jessie L. and {L{\'o}pez-Morales}, Mercedes and
         {Swain}, Mark R. and {Zellem}, Robert T. and {Ballard}, Sarah and
         {Barclay}, Thomas and {Barstow}, Joanna K. and {Batalha}, Natasha E. and
         {Beatty}, Thomas G. and {Berta-Thompson}, Zach and {Birkby}, Jayne and
         {Buchhave}, Lars A. and {Charbonneau}, David and {Cowan}, Nicolas B. and
         {Crossfield}, Ian and {de Val-Borro}, Miguel and {Doyon}, Ren{\'e} and
         {Dragomir}, Diana and {Gaidos}, Eric and {Heng}, Kevin and {Hu}, Renyu and
         {Kane}, Stephen R. and {Kreidberg}, Laura and {Mallonn}, Matthias and
         {Morley}, Caroline V. and {Narita}, Norio and {Nascimbeni}, Valerio and
         {Pall{\'e}}, Enric and {Quintana}, Elisa V. and {Rauscher}, Emily and
         {Seager}, Sara and {Shkolnik}, Evgenya L. and {Sing}, David K. and
         {Sozzetti}, Alessandro and {Stassun}, Keivan G. and {Valenti}, Jeff A. and
         {von Essen}, Carolina},
        title = "{A Framework for Prioritizing the TESS Planetary Candidates Most Amenable to Atmospheric Characterization}",
      journal = {\pasp},
         year = 2018,
        month = nov,
       volume = {130},
       number = {993},
        pages = {114401},
          doi = {10.1088/1538-3873/aadf6f},
archivePrefix = {arXiv},
       eprint = {1805.03671},
 primaryClass = {astro-ph.EP},
       adsurl = {https://ui.adsabs.harvard.edu/abs/2018PASP..130k4401K}
}

@ARTICLE{Ricker2015,
   author = {{Ricker}, G.~R. and {Winn}, J.~N. and {Vanderspek}, R. and {Latham}, D.~W. and 
	{Bakos}, G.~{\'A}. and {Bean}, J.~L. and {Berta-Thompson}, Z.~K. and 
	{Brown}, T.~M. and {Buchhave}, L. and {Butler}, N.~R. and {Butler}, R.~P. and 
	{Chaplin}, W.~J. and {Charbonneau}, D. and {Christensen-Dalsgaard}, J. and 
	{Clampin}, M. and {Deming}, D. and {Doty}, J. and {De Lee}, N. and 
	{Dressing}, C. and {Dunham}, E.~W. and {Endl}, M. and {Fressin}, F. and 
	{Ge}, J. and {Henning}, T. and {Holman}, M.~J. and {Howard}, A.~W. and 
	{Ida}, S. and {Jenkins}, J.~M. and {Jernigan}, G. and {Johnson}, J.~A. and 
	{Kaltenegger}, L. and {Kawai}, N. and {Kjeldsen}, H. and {Laughlin}, G. and 
	{Levine}, A.~M. and {Lin}, D. and {Lissauer}, J.~J. and {MacQueen}, P. and 
	{Marcy}, G. and {McCullough}, P.~R. and {Morton}, T.~D. and 
	{Narita}, N. and {Paegert}, M. and {Palle}, E. and {Pepe}, F. and 
	{Pepper}, J. and {Quirrenbach}, A. and {Rinehart}, S.~A. and 
	{Sasselov}, D. and {Sato}, B. and {Seager}, S. and {Sozzetti}, A. and 
	{Stassun}, K.~G. and {Sullivan}, P. and {Szentgyorgyi}, A. and 
	{Torres}, G. and {Udry}, S. and {Villasenor}, J.},
    title = "{Transiting Exoplanet Survey Satellite (TESS)}",
  journal = {Journal of Astronomical Telescopes, Instruments, and Systems},
     year = 2015,
    month = jan,
   volume = 1,
   number = 1,
      eid = {014003},
    pages = {014003},
      doi = {10.1117/1.JATIS.1.1.014003},
   adsurl = {http://adsabs.harvard.edu/abs/2015JATIS...1a4003R}
}

@ARTICLE{Rackham2023,
       author = {{Rackham}, Benjamin V. and {Espinoza}, N{\'e}stor and {Berdyugina}, Svetlana V. and {Korhonen}, Heidi and {MacDonald}, Ryan J. and {Montet}, Benjamin T. and {Morris}, Brett M. and {Oshagh}, Mahmoudreza and {Shapiro}, Alexander I. and {Unruh}, Yvonne C. and {Quintana}, Elisa V. and {Zellem}, Robert T. and {Apai}, D{\'a}niel and {Barclay}, Thomas and {Barstow}, Joanna K. and {Bruno}, Giovanni and {Carone}, Ludmila and {Casewell}, Sarah L. and {Cegla}, Heather M. and {Criscuoli}, Serena and {Fischer}, Catherine and {Fournier}, Damien and {Giampapa}, Mark S. and {Giles}, Helen and {Iyer}, Aishwarya and {Kopp}, Greg and {Kostogryz}, Nadiia M. and {Krivova}, Natalie and {Mallonn}, Matthias and {McGruder}, Chima and {Molaverdikhani}, Karan and {Newton}, Elisabeth R. and {Panja}, Mayukh and {Peacock}, Sarah and {Reardon}, Kevin and {Roettenbacher}, Rachael M. and {Scandariato}, Gaetano and {Solanki}, Sami and {Stassun}, Keivan G. and {Steiner}, Oskar and {Stevenson}, Kevin B. and {Tregloan-Reed}, Jeremy and {Valio}, Adriana and {Wedemeyer}, Sven and {Welbanks}, Luis and {Yu}, Jie and {Alam}, Munazza K. and {Davenport}, James R.~A. and {Deming}, Drake and {Dong}, Chuanfei and {Ducrot}, Elsa and {Fisher}, Chloe and {Gilbert}, Emily and {Kostov}, Veselin and {L{\'o}pez-Morales}, Mercedes and {Line}, Mike and {Mo{\v{c}}nik}, Teo and {Mullally}, Susan and {Paudel}, Rishi R. and {Ribas}, Ignasi and {Valenti}, Jeff A.},
        title = "{The effect of stellar contamination on low-resolution transmission spectroscopy: needs identified by NASA's Exoplanet Exploration Program Study Analysis Group 21}",
      journal = {RAS Techniques and Instruments},
         year = 2023,
        month = jan,
       volume = {2},
       number = {1},
        pages = {148-206},
          doi = {10.1093/rasti/rzad009},
archivePrefix = {arXiv},
       eprint = {2201.09905},
 primaryClass = {astro-ph.IM},
       adsurl = {https://ui.adsabs.harvard.edu/abs/2023RASTI...2..148R}
}

@ARTICLE{Foote2023,
       author = {{Foote}, Trevor O. and {Barclay}, Thomas and {Hedges}, Christina L. and {Lewis}, Nikole K. and {Quintana}, Elisa V. and {Rackham}, Benjamin V. and {Col{\'o}n}, Knicole and {Ciardi}, David},
        title = "{Schedule optimization for transiting exoplanet observations with National Aeronautics and Space Administration's Pandora SmallSat mission}",
      journal = {Journal of Astronomical Telescopes, Instruments, and Systems},
         year = 2023,
        month = oct,
       volume = {9},
          eid = {047001},
        pages = {047001},
          doi = {10.1117/1.JATIS.9.4.047001},
archivePrefix = {arXiv},
       eprint = {2305.02285},
 primaryClass = {astro-ph.IM},
       adsurl = {https://ui.adsabs.harvard.edu/abs/2023JATIS...9d7001F}
}

@ARTICLE{Wolszczan1992,
       author = {{Wolszczan}, A. and {Frail}, D.~A.},
        title = "{A planetary system around the millisecond pulsar PSR1257 + 12}",
      journal = {\nat},
         year = 1992,
        month = jan,
       volume = {355},
       number = {6356},
        pages = {145-147},
          doi = {10.1038/355145a0},
       adsurl = {https://ui.adsabs.harvard.edu/abs/1992Natur.355..145W}
}

@ARTICLE{Mayor1995,
       author = {{Mayor}, Michel and {Queloz}, Didier},
        title = "{A Jupiter-mass companion to a solar-type star}",
      journal = {\nat},
         year = 1995,
        month = nov,
       volume = {378},
       number = {6555},
        pages = {355-359},
          doi = {10.1038/378355a0},
       adsurl = {https://ui.adsabs.harvard.edu/abs/1995Natur.378..355M}
}

@ARTICLE{Christiansen2025,
       author = {{Christiansen}, Jessie L. and {McElroy}, Douglas L. and {Harbut}, Marcy and {Ciardi}, David R. and {Crane}, Megan and {Good}, John and {Hardegree-Ullman}, Kevin K. and {Kesseli}, Aurora Y. and {Lund}, Michael B. and {Lynn}, Meca and {Muthiar}, Ananda and {Nilsson}, Ricky and {Oluyide}, Toba and {Papin}, Michael and {Rivera}, Amalia and {Swain}, Melanie and {Susemiehl}, Nicholas D. and {Tam}, Raymond and {van Eyken}, Julian and {Beichman}, Charles},
        title = "{The NASA Exoplanet Archive and Exoplanet Follow-up Observing Program: Data, Tools, and Usage}",
      journal = {\psj},
         year = 2025,
        month = aug,
       volume = {6},
       number = {8},
          eid = {186},
        pages = {186},
          doi = {10.3847/PSJ/ade3c2},
archivePrefix = {arXiv},
       eprint = {2506.03299},
 primaryClass = {astro-ph.EP},
       adsurl = {https://ui.adsabs.harvard.edu/abs/2025PSJ.....6..186C}
}

@ARTICLE{Chen2017,
       author = {{Chen}, Jingjing and {Kipping}, David},
        title = "{Probabilistic Forecasting of the Masses and Radii of Other Worlds}",
      journal = {\apj},
         year = 2017,
        month = jan,
       volume = {834},
       number = {1},
          eid = {17},
        pages = {17},
          doi = {10.3847/1538-4357/834/1/17},
archivePrefix = {arXiv},
       eprint = {1603.08614},
 primaryClass = {astro-ph.EP},
       adsurl = {https://ui.adsabs.harvard.edu/abs/2017ApJ...834...17C}
}

@ARTICLE{Mordasini2024,
       author = {{Mordasini}, Christoph and {Burn}, Remo},
        title = "{Planet Formation{\textemdash}Observational Constraints, Physical Processes, and Compositional Patterns}",
      journal = {Reviews in Mineralogy and Geochemistry},
         year = 2024,
        month = jul,
       volume = {90},
       number = {1},
        pages = {55-112},
          doi = {10.2138/rmg.2024.90.03},
archivePrefix = {arXiv},
       eprint = {2404.15555},
 primaryClass = {astro-ph.EP},
       adsurl = {https://ui.adsabs.harvard.edu/abs/2024RvMG...90...55M}
}

@ARTICLE{Vallenari2023,
       author = {{Gaia Collaboration} and {Vallenari}, A. and {Brown}, A.~G.~A. and {Prusti}, T. and {de Bruijne}, J.~H.~J. and {Arenou}, F. and {Babusiaux}, C. and {Biermann}, M. and {Creevey}, O.~L. and {Ducourant}, C. and {Evans}, D.~W. and {Eyer}, L. and {Guerra}, R. and {Hutton}, A. and {Jordi}, C. and {Klioner}, S.~A. and {Lammers}, U.~L. and {Lindegren}, L. and {Luri}, X. and {Mignard}, F. and {Panem}, C. and {Pourbaix}, D. and {Randich}, S. and {Sartoretti}, P. and {Soubiran}, C. and {Tanga}, P. and {Walton}, N.~A. and {Bailer-Jones}, C.~A.~L. and {Bastian}, U. and {Drimmel}, R. and {Jansen}, F. and {Katz}, D. and {Lattanzi}, M.~G. and {van Leeuwen}, F. and {Bakker}, J. and {Cacciari}, C. and {Casta{\~n}eda}, J. and {De Angeli}, F. and {Fabricius}, C. and {Fouesneau}, M. and {Fr{\'e}mat}, Y. and {Galluccio}, L. and {Guerrier}, A. and {Heiter}, U. and {Masana}, E. and {Messineo}, R. and {Mowlavi}, N. and {Nicolas}, C. and {Nienartowicz}, K. and {Pailler}, F. and {Panuzzo}, P. and {Riclet}, F. and {Roux}, W. and {Seabroke}, G.~M. and {Sordo}, R. and {Th{\'e}venin}, F. and {Gracia-Abril}, G. and {Portell}, J. and {Teyssier}, D. and {Altmann}, M. and {Andrae}, R. and {Audard}, M. and {Bellas-Velidis}, I. and {Benson}, K. and {Berthier}, J. and {Blomme}, R. and {Burgess}, P.~W. and {Busonero}, D. and {Busso}, G. and {C{\'a}novas}, H. and {Carry}, B. and {Cellino}, A. and {Cheek}, N. and {Clementini}, G. and {Damerdji}, Y. and {Davidson}, M. and {de Teodoro}, P. and {Nu{\~n}ez Campos}, M. and {Delchambre}, L. and {Dell'Oro}, A. and {Esquej}, P. and {Fern{\'a}ndez-Hern{\'a}ndez}, J. and {Fraile}, E. and {Garabato}, D. and {Garc{\'\i}a-Lario}, P. and {Gosset}, E. and {Haigron}, R. and {Halbwachs}, J.-L. and {Hambly}, N.~C. and {Harrison}, D.~L. and {Hern{\'a}ndez}, J. and {Hestroffer}, D. and {Hodgkin}, S.~T. and {Holl}, B. and {Jan{\ss}en}, K. and {Jevardat de Fombelle}, G. and {Jordan}, S. and {Krone-Martins}, A. and {Lanzafame}, A.~C. and {L{\"o}ffler}, W. and {Marchal}, O. and {Marrese}, P.~M. and {Moitinho}, A. and {Muinonen}, K. and {Osborne}, P. and {Pancino}, E. and {Pauwels}, T. and {Recio-Blanco}, A. and {Reyl{\'e}}, C. and {Riello}, M. and {Rimoldini}, L. and {Roegiers}, T. and {Rybizki}, J. and {Sarro}, L.~M. and {Siopis}, C. and {Smith}, M. and {Sozzetti}, A. and {Utrilla}, E. and {van Leeuwen}, M. and {Abbas}, U. and {{\'A}brah{\'a}m}, P. and {Abreu Aramburu}, A. and {Aerts}, C. and {Aguado}, J.~J. and {Ajaj}, M. and {Aldea-Montero}, F. and {Altavilla}, G. and {{\'A}lvarez}, M.~A. and {Alves}, J. and {Anders}, F. and {Anderson}, R.~I. and {Anglada Varela}, E. and {Antoja}, T. and {Baines}, D. and {Baker}, S.~G. and {Balaguer-N{\'u}{\~n}ez}, L. and {Balbinot}, E. and {Balog}, Z. and {Barache}, C. and {Barbato}, D. and {Barros}, M. and {Barstow}, M.~A. and {Bartolom{\'e}}, S. and {Bassilana}, J.-L. and {Bauchet}, N. and {Becciani}, U. and {Bellazzini}, M. and {Berihuete}, A. and {Bernet}, M. and {Bertone}, S. and {Bianchi}, L. and {Binnenfeld}, A. and {Blanco-Cuaresma}, S. and {Blazere}, A. and {Boch}, T. and {Bombrun}, A. and {Bossini}, D. and {Bouquillon}, S. and {Bragaglia}, A. and {Bramante}, L. and {Breedt}, E. and {Bressan}, A. and {Brouillet}, N. and {Brugaletta}, E. and {Bucciarelli}, B. and {Burlacu}, A. and {Butkevich}, A.~G. and {Buzzi}, R. and {Caffau}, E. and {Cancelliere}, R. and {Cantat-Gaudin}, T. and {Carballo}, R. and {Carlucci}, T. and {Carnerero}, M.~I. and {Carrasco}, J.~M. and {Casamiquela}, L. and {Castellani}, M. and {Castro-Ginard}, A. and {Chaoul}, L. and {Charlot}, P. and {Chemin}, L. and {Chiaramida}, V. and {Chiavassa}, A. and {Chornay}, N. and {Comoretto}, G. and {Contursi}, G. and {Cooper}, W.~J. and {Cornez}, T. and {Cowell}, S. and {Crifo}, F. and {Cropper}, M. and {Crosta}, M. and {Crowley}, C. and {Dafonte}, C. and {Dapergolas}, A. and {David}, M. and {David}, P. and {de Laverny}, P. and {De Luise}, F. and {De March}, R.},
        title = "{Gaia Data Release 3. Summary of the content and survey properties}",
      journal = {\aap},
         year = 2023,
        month = jun,
       volume = {674},
          eid = {A1},
        pages = {A1},
          doi = {10.1051/0004-6361/202243940},
archivePrefix = {arXiv},
       eprint = {2208.00211},
 primaryClass = {astro-ph.GA},
       adsurl = {https://ui.adsabs.harvard.edu/abs/2023A&A...674A...1G}
}

@ARTICLE{Moran2023,
       author = {{Moran}, Sarah E. and {Stevenson}, Kevin B. and {Sing}, David K. and {MacDonald}, Ryan J. and {Kirk}, James and {Lustig-Yaeger}, Jacob and {Peacock}, Sarah and {Mayorga}, L.~C. and {Bennett}, Katherine A. and {L{\'o}pez-Morales}, Mercedes and {May}, E.~M. and {Rustamkulov}, Zafar and {Valenti}, Jeff A. and {Adams Redai}, J{\'e}a I. and {Alam}, Munazza K. and {Batalha}, Natasha E. and {Fu}, Guangwei and {Gonzalez-Quiles}, Junellie and {Highland}, Alicia N. and {Kruse}, Ethan and {Lothringer}, Joshua D. and {Ortiz Ceballos}, Kevin N. and {Sotzen}, Kristin S. and {Wakeford}, Hannah R.},
        title = "{High Tide or Riptide on the Cosmic Shoreline? A Water-rich Atmosphere or Stellar Contamination for the Warm Super-Earth GJ 486b from JWST Observations}",
      journal = {\apjl},
         year = 2023,
        month = may,
       volume = {948},
       number = {1},
          eid = {L11},
        pages = {L11},
          doi = {10.3847/2041-8213/accb9c},
archivePrefix = {arXiv},
       eprint = {2305.00868},
 primaryClass = {astro-ph.EP},
       adsurl = {https://ui.adsabs.harvard.edu/abs/2023ApJ...948L..11M}
}

@ARTICLE{Canas2026,
       author = {{Ca{\~n}as}, Caleb I. and {Lustig-Yaeger}, Jacob and {Tsai}, Shang-Min and {M{\"u}ller}, Simon and {Helled}, Ravit and {Kanodia}, Shubham and {Louie}, Dana R. and {Guzm{\'a}n Caloca}, Giannina and {Gao}, Peter and {Libby-Roberts}, Jessica and {Hardegree-Ullman}, Kevin K. and {Col{\'o}n}, Knicole D. and {Czekala}, Ian and {Delamer}, Megan and {Han}, Te and {Lin}, Andrea S.~J. and {Mahadevan}, Suvrath and {May}, Erin M. and {Ninan}, Joe P. and {Piette}, Anjali A.~A. and {Stef{\'a}nsson}, Gu{\dh}mundur and {Stevenson}, Kevin B. and {Teske}, Johanna and {Wallack}, Nicole L.},
        title = "{GEMS JWST: Transmission Spectroscopy of TOI-5205b Reveals Significant Stellar Contamination and a Metal-poor Atmosphere}",
      journal = {\aj},
         year = 2026,
        month = apr,
       volume = {171},
       number = {4},
          eid = {260},
        pages = {260},
          doi = {10.3847/1538-3881/ae4976},
archivePrefix = {arXiv},
       eprint = {2502.06966},
 primaryClass = {astro-ph.EP},
       adsurl = {https://ui.adsabs.harvard.edu/abs/2026AJ....171..260C}
}

@INPROCEEDINGS{Quintana2024,
       author = {{Quintana}, Elisa V. and {Dotson}, Jessie L. and {Col{\'o}n}, Knicole D. and {Barclay}, Thomas and {Supsinskas}, Pete and {Karburn}, Jordan and {Apai}, D{\'a}niel and {Hedges}, Christina and {Rackham}, Benjamin V. and {Rowe}, Jason F. and {Allen}, Natalie H. and {Bonney}, Paul and {Cano}, Samuel and {Christiansen}, Jessie L. and {Ciardi}, David and {Espinoza}, N{\'e}stor and {Foote}, Trevor O. and {Gilbert}, Emily A. and {Greene}, Thomas P. and {Hoffman}, Kelsey and {Hord}, Benjamin J. and {Iyer}, Aishwarya and {Kesseli}, Aurora and {Kostov}, Veselin B. and {Lewis}, Nikole K. and {Logsdon}, Sarah E. and {Mann}, Andrew W. and {Mansfield}, Megan and {Mason}, James and {Morris}, Brett M. and {Mosby}, Gregory and {Mullally}, Susan E. and {Newton}, Elisabeth R. and {Nguyen}, Fuda and {Schlieder}, Joshua E. and {Stevenson}, Kevin B. and {Wiser}, Lindsey S. and {Youngblood}, Allison and {Zellem}, Robert T.},
        title = "{The Pandora SmallSat: multiwavelength characterization of exoplanets and their host stars}",
    booktitle = {Space Telescopes and Instrumentation 2024: Optical, Infrared, and Millimeter Wave},
         year = 2024,
       editor = {{Coyle}, Laura E. and {Matsuura}, Shuji and {Perrin}, Marshall D.},
       series = {Society of Photo-Optical Instrumentation Engineers (SPIE) Conference Series},
       volume = {13092},
        month = aug,
          eid = {1309214},
        pages = {1309214},
          doi = {10.1117/12.3020633},
       adsurl = {https://ui.adsabs.harvard.edu/abs/2024SPIE13092E..14Q}
}

@ARTICLE{Charbonneau2000,
       author = {{Charbonneau}, David and {Brown}, Timothy M. and {Latham}, David W. and {Mayor}, Michel},
        title = "{Detection of Planetary Transits Across a Sun-like Star}",
      journal = {\apjl},
         year = 2000,
        month = jan,
       volume = {529},
       number = {1},
        pages = {L45-L48},
          doi = {10.1086/312457},
archivePrefix = {arXiv},
       eprint = {astro-ph/9911436},
 primaryClass = {astro-ph},
       adsurl = {https://ui.adsabs.harvard.edu/abs/2000ApJ...529L..45C}
}

@ARTICLE{Charbonneau2002,
       author = {{Charbonneau}, David and {Brown}, Timothy M. and {Noyes}, Robert W. and {Gilliland}, Ronald L.},
        title = "{Detection of an Extrasolar Planet Atmosphere}",
      journal = {\apj},
         year = 2002,
        month = mar,
       volume = {568},
       number = {1},
        pages = {377-384},
          doi = {10.1086/338770},
archivePrefix = {arXiv},
       eprint = {astro-ph/0111544},
 primaryClass = {astro-ph},
       adsurl = {https://ui.adsabs.harvard.edu/abs/2002ApJ...568..377C}
}
\bibliographystyle{spiebib} 

\end{document}